\documentclass[11pt]{elsart}
\usepackage{amsmath,amssymb}
\usepackage{epsfig}
\usepackage{algorithm}
\usepackage{algorithmic}
\newcommand{\bv}{\boldsymbol{v}}
\newcommand{\bV}{\boldsymbol{V}}
\newcommand{\bkhat}{\boldsymbol{\hat k}}
\newcommand{\bx}{\boldsymbol{x}}
\newcommand{\bF}{\boldsymbol{F}}
\newcommand{\bj}{\boldsymbol{j}}
\newcommand{\bn}{\boldsymbol{n}}
\begin{document}

\begin{frontmatter}

\title{Solving Kinetic Equations on GPUs I: Model Kinetic Equations} 

\author{A. Frezzotti},
\ead{aldo.frezzotti@polimi.it}
\author{G. P. Ghiroldi},
\ead{gian.ghiroldi@mail.polimi.it}
\author{L. Gibelli\corauthref{cor}}
\corauth[cor]{Corresponding author.}
\ead{livio.gibelli@polimi.it}

\address{Politecnico di Milano, 
         Dipartimento di Matematica,
         Piazza Leonardo da Vinci 32,
         20133 Milano, Italy}

\begin{abstract}
We present an algorithm specifically tailored for solving kinetic 
equations onto GPUs. The efficiency of the
algorithm is demonstrated by solving the one-dimensional shock wave 
structure problem and a two-dimensional low Mach number driven cavity flow. 
Computational results show that it is possible to cut 
down the computing time of the sequential codes of two order of magnitudes. 
The algorithm can easily be extended to three-dimensional flows and more general collision models. 
\end{abstract}


\end{frontmatter}

\section{Introduction}

A recent trend emerging in computational physics stems from the 
availability of low cost general purpose graphics processing units (GPUs).
GPUs have been used to accelerate CPU critical applications such as 
simulations of hypersonic flows \cite{eld08}, magnetized plasma \cite{sdg08} 
and molecular dynamics \cite{alt08}. 
However, no applications to kinetic theory of gases seem to have been 
considered yet. Kinetic theory of gases deals with 
non-equilibrium gas flows which are met in several different physical 
situations ranging from the re-entry of spacecraft in upper planetary atmospheres to fluid-structure interaction in small-scale devices \cite{g99,lgffc07}. 
The dynamics of dilute 
(or rarefied) gas flows  is governed by the
Boltzmann equation \cite{c88} which takes the form
\begin{eqnarray}
&&\frac{\partial f }{\partial t}+\bv\circ\nabla_{\bx}f+\frac{1}{m}\nabla_{\bv}\circ(\bF f)=\mathcal{C}(f,f) \label{eq:BE}\\
&&\mathcal{C}(f,f)=\int \left[f(\bx,\bv^*|t)f(\bx,\bv_{1}^*|t)-\right. \nonumber \\ 
&&\left. 
-f(\bx,\bv|t)f(\bx,\bv_{1}|t)\right]\sigma(\|\bv_r\|,\bkhat\circ \bv_r)\|\bv_r\| d\bv_1 d^2\bkhat
\label{eq:Collint}
\end{eqnarray}
when written for a gas composed by a single monatomic species whose atoms have mass $m$.
In Eqs. (\ref{eq:BE},\ref{eq:Collint}), $f(\bx,\bv|t)$ denotes the distribution function of atomic velocities $\bv$ at spatial location $\bx$ and time $t$, $\bF(\bx,\bv|t)$ is an assigned external force field, whereas $\mathcal{C}(f,f)$  gives the collisional rate of change of $f$ at the phase space point $(\bx,\bv)$ at time $t$. As is clear from Eq. (\ref{eq:Collint}), $\mathcal{C}(f,f)$ is a non-linear functional of $f$, whose precise structure depends on the assumed atomic interaction forces through the differential cross section $\sigma(\|\bv_r\|,\bkhat\circ \bv_r)$. The dynamics of binary encounters determines $\sigma$ as a function of the modulus $\|\bv_r\|$ of the relative velocity $\bv_r=\bv_1-\bv$ of two colliding atoms and of the orientation of the unit impact vector $\bkhat$ with respect to $\bv_r$ \cite{cc90}.
The collisional dynamics also determines the pre-collisional velocities $\bv^*$ and $\bv_{1}^*$
which are changed into $\bv$ and $\bv_{1}$ by a binary collision. For illustration purposes, it is worth mentioning that the collision integral $\mathcal{C}(f,f)$ simplifies to
\begin{equation}
\mathcal{C}(f,f)=\frac{d^2}{2}\int \left[f(\bx,\bv^*|t)f(\bx,\bv_{1}^*|t)
-f(\bx,\bv|t)f(\bx,\bv_{1}|t)\right]|\bkhat\circ \bv_r| d\bv_1 d^2\bkhat
\end{equation}
for a dilute gas of hard spheres of diameter $d$. In this case  $\bv^*$ and $\bv_{1}^*$
are obtained from $\bv$, $\bv_{1}$ and $\bkhat$ by the simple relationships
\begin{eqnarray}
\bv^*&=& \bv+(\bv_r\circ \bkhat)\bkhat \\
\bv_{1}^*&=& \bv_{1}-(\bv_r\circ \bkhat)\bkhat
\end{eqnarray}
Obtaining numerical solutions of the Boltzmann equation for realistic flow conditions is a challenging task because the unknown function depends, in principle, on seven variables.
Moreover, the computation of $\mathcal{C}(f,f)$ requires the approximate evaluation of a fivefold integral. Numerical methods for rarefied gas dynamics studies can be roughly divided into
three groups:
\begin{itemize}
 \item[(a)] Particle methods
\item[(b)] Semi-regular methods
\item[(c)] Regular methods  
\end{itemize}
Methods in group (a) originate from the Direct Simulation Monte Carlo (DSMC) scheme proposed by G.A. Bird \cite{b94}. They are by far the most popular and widely used simulation methods in rarefied gas dynamics. The distribution function is represented by a number of mathematical particles which move in the computational domain and collide according to stochastic rules derived from Eqs. (\ref{eq:BE},\ref{eq:Collint}). Macroscopic flow properties are usually obtained by time averaging particle properties. If the averaging time is long enough, then accurate flow simulations can be obtained by a relatively small number of particles. The method can be easily extended to deal with mixtures of chemically reacting polyatomic species \cite{b94} and to dense fluids \cite{fgl05}. Although DSMC (in its traditional implementation) is to be recommended in simulating most of rarefied gas flows, it is not well suited to the simulation of low Mach number or unsteady flows. Attempts have been made to extend DSMC in order to improve its capability to capture the small deviations from the equilibrium condition met in low Mach number flows \cite{hh07,w08}. 
However, in simulating high frequency unsteady flows, typical of microfluidics application to MEMS \cite{g99}, the possibility of time averaging is lost or reduced. Acceptable accuracy can then be achieved by increasing the number of simulation particles or superposing several flow snapshots
obtained from statistically independent simulations of the same flow; in both cases the computing effort is considerably increased.\\
Methods in groups (b) and (c) adopt similar strategies in discretizing the distribution function on a regular grid in the phase space and in using finite difference schemes to approximate the streaming term on the l.h.s of Eq. (\ref{eq:BE}). However, they differ in the way the collision integral is evaluated.
In semi-regular methods $\mathcal{C}(f,f)$ is computed by Monte Carlo or quasi Monte Carlo quadrature methods \cite{f91,t05} whereas deterministic integration schemes are used in regular methods \cite{a01}. Whatever method is chosen to compute the collision term, the adoption of a grid in the phase space considerably limits the applicability of methods (b) and (c) to problems where particular symmetries reduce the number of spatial and velocity variables. As a matter of fact, a spatially three-dimensional problem would require a memory demanding six-dimensional phase space grid. Extensions to polyatomic gases are possible 
\cite{f07} but the necessity to store additional variables associated with internal degrees of freedom further limits the applications to multi-dimensional flows.     
In spite of the drawbacks listed above, the direct solution of the Boltzmann equation by
semi-regular or regular methods is a valid alternative to particle schemes in studying unsteady or low speed flows. Actually, when the deviation from equilibrium is small a limited number of grid points in the velocity space is sufficient to provide accurate and noise free approximations of $f$, therefore simulations of multi-dimensional flows are feasible on modern personal computers in a wide range of Knudsen numbers \cite{VVS08}.\\
An important feature of kinetic equations for dilute gases is the locality of the collision term;
the collisional rate of change $\mathcal{C}(f,f)$ at the spatial location $\bx$ is completely determined by $f(\bx,\bv|t)$. Hence, the time consuming evaluation of the collision integral can 
be concurrently executed at each spatial grid point on parallel computers.
As shown below, the numerical algorithm associated with regular or semi-regular methods is ideally suited for the parallel architecture provided by commercially available GPUs. 
The aim of the paper is to describe an efficient algorithm specifically
tailored for solving kinetic equations onto GPUs using 
$\mbox{CUDA}^{\mbox{\tiny TM}}$  
programming model \cite{n08}. The efficiency of the algorithm is assessed by solving the classical one-dimensional shock wave structure and a low speed 
two-dimensional driven cavity flow.
It is shown that it is possible to cut the computing time of the sequential 
codes of two order of magnitudes by a proper reformulation of the algorithm to be executed on a GPU. \\
In order to make the algorithm development easier, the computations presented here have been performed by replacing $\mathcal{C}(f,f)$ with its simpler BGKW approximation \cite{c88}.
This choice eliminates the intricacies connected with the numerical evaluation of the Boltzmann
collision integral and allows easier identification of bottlenecks and optimization strategies.
As will be shown in a separate paper, the full Boltzmann equation can be solved within the same
general algorithmic framework by adopting a Monte Carlo quadrature method.\\ 
This paper is organized as follows. Section II is devoted to a concise description of the mathematical model and the adopted numerical method. In Section III the key aspects of the GPU
hardware architecture and $\mbox{CUDA}^{\mbox{\tiny TM}}$  
programming language are briefly described. Sections IV and V
are devoted to the description of the test problems and the discussion of the results.
Concluding remarks are presented in Section VI.

\section{Theoretical and numerical background}\label{sec:TNB}
Both from the theoretical and computational point of view, it is often convenient
to replace the full Boltzmann equation with a model equation having a simplified
collision term. In the kinetic model proposed by Bhatnagar Gross and Krook \cite{bgk54} and
independently by Welander \cite{w54}, $\mathcal{C}(f,f)$ 
is replaced by the expression
$\nu\left(\Phi-f\right)$. Accordingly, Eq. (\ref{eq:BE}) is turned into the following
kinetic equation:
\begin{equation}
\frac{\partial f }{\partial t}+\bv\circ\nabla_{\bx}f+\frac{1}{m}\nabla_{\bv}\circ(\bF f)=\nu\left(\Phi-f\right)
\label{eq:BGKW}
\end{equation} 
In Eq. (\ref{eq:BGKW}) $\nu$ is the collision frequency, whereas $\Phi$ is the local equilibrium Maxwellian distribution function given
by the expression
\begin{equation}
 \Phi(\bx,\bv|t)=\frac{n(\bx|t)}{\left[2\pi R T(\bx|t) \right]^{3/2}}\exp\left\{-\frac{[\bv-\bV(\bx|t)]^2}{2RT(\bx|t)}\right\}
\label{eq:Maxw}
\end{equation}
If $\nu$ does not depend on the velocity $\bv$, then conservation of mass, momentum and energy requires that $n$, $\bV$ and $T$ in Eq. (\ref{eq:Maxw}) coincide with the local values of density, bulk
velocity and temperature obtained from $f$ by the relationships
 \begin{equation}
\label{macroscopic}
n= \int f d{\bv} \mbox{\hspace{1cm}}
{\bf V} = \frac{1}{n} \int f {\bv} d{\bv} \mbox{\hspace{1cm}}
T = \frac{1}{3Rn} \int f ({\bv}-{\bf V})^{2} d{\bv}
\end{equation}
being $R$ the specific gas constant. The above expressions show that Eq. (\ref{eq:BGKW}) is a strongly non-linear integro-differential equation, in spite of the linear appearance of its r.h.s..\\
As is well known, the BGKW model predicts an incorrect value of the Prandtl number in the hydrodynamic limit \cite{c88}.
Hence, $\nu$ can be adjusted to obtain either the correct viscosity or heat conductivity, but not both. If viscosity is selected, then $\nu$ is given by the following expression:
 \begin{equation}
\nu = \frac{n RT}{\mu}
\label{eq:frequency}
\end{equation}
being $\mu(T)$ the gas viscosity.

\subsection{Outline of the numerical method}
\label{sec:TBN}
In view of the exploratory nature of the present work, Eq. (\ref{eq:BGKW}) has been solved by a simple numerical method which will be illustrated on a spatially one-dimensional problem. The extension to two or three-dimensional geometries is straightforward.\\
In absence of external forces and in one-dimensional slab geometry Eq. (\ref{eq:BGKW}) takes the form:
\begin{equation}
\frac{\partial f }{\partial t}+v_x\frac{\partial f }{\partial x}=\nu\left(\Phi-f\right)
\label{eq:1D}
\end{equation} 
where $x$ is the single spatial coordinate and $v_x$ the $x$-component of the velocity vector $\bv=(v_x,v_y,v_z)$.
The spatial domain is a finite interval of the real axis, divided into $N_x$ cells of equal size $\Delta x$.
The infinite three-dimensional velocity space is replaced by a rectangular box divided into $N_v=N_{v_x}\times N_{v_y}\times N_{v_z}$
cells of equal volume $\Delta \mathcal{V}$, $N_{v_\alpha}$ being the number of velocity nodes associated with the velocity component
$v_\alpha$. The size and position of the ``velocity box'' in the velocity space have to be properly chosen, in order to contain the
significant part of $f$ at any spatial position. The distribution function is assumed to be constant within each cell of the phase space. Hence, $f$ is represented by the array $f_{i,\bj}(t)=f(x(i),v_x(j_x),v_y(j_y),v_z(j_z)|t)$, being $x(i),v_x(j_x),v_y(j_y),v_z(j_z)$ the values of the spatial coordinate and velocity components in the center of the phase space cell
$(i,{\bf j})$ and $\bj=(j_x,j_y,j_z)$.\\
The algorithm that advances $f_{i,\bj}(t)$ to $f_{i,\bj}(t+\Delta t)$ is constructed by time-splitting the evolution operator into a free streaming step, in which the r.h.s.
of Eq. (\ref{eq:BGKW}) is neglected, and a purely collisional step, in which spatial motion
is frozen and only the effect of the r.h.s. are taken into account. More precisely, the distribution function $f_{i,\bj}^n=f_{i,\bj}(t_n)$ at time level $t_n$ is advanced to its 
value $f_{i,\bj}^{n+1}=f_{i,\bj}(t_{n+1})$ at time level $t_{n+1}=t_n+\Delta t$ by computing an intermediate value $\tilde f_{i,\bj}^{n+1}$ from the free streaming equation
\begin{equation}
\frac{\partial f }{\partial t}+v_x\frac{\partial f }{\partial x}=0
\label{eq:freestreaming}
\end{equation} 
Eq. (\ref{eq:freestreaming}) is solved by a simple first order upwind scheme
\begin{equation}
\tilde f_{i,\bj}^{n+1}=
\begin{cases}
\displaystyle{\left(1-\frac{v_x(j_x)\Delta t}{\Delta x}\right)f_{i,\bj}^{n}+\frac{v_x(j_x)\Delta t}{\Delta x}f_{i-1,\bj}^{n}} & v_x(j_x)>0 \\
\displaystyle{\left(1+\frac{v_x(j_x)\Delta t}{\Delta x}\right)f_{i,\bj}^{n}-\frac{v_x(j_x)\Delta t}{\Delta x}f_{i+1,\bj}^{n}} & v_x(j_x)<0
\end{cases}
\label{eq:upwind}
\end{equation}
After completing the free streaming step, macroscopic variables $n_i$, $\bV_i$ and $T_i$ are computed at each spatial grid point and  $f_{i,\bj}^{n+1}$ is finally obtained by solving the
homogeneous relaxation equation 
\begin{equation}
 \frac{\partial f}{\partial t}=\nu(\Phi-f)
\label{eq:homrel}
\end{equation}
Since $n$, $\bV$ and $T$ are conserved during homogeneous relaxation, Eq. (\ref{eq:homrel})
can be exactly solved to obtain
\begin{equation}
\label{relaxation}
 f_{i,\bj}^{n+1}=\left[1-\exp(-\nu_i\Delta t)\right]\Phi_{i,\bj}+\exp(-\nu_i\Delta t)\tilde f_{i,\bj}^{n+1}
\end{equation}
in each cell $(i,\bj)$ of the phase space.
The time step $\Delta t$ has been set equal to a fraction of $1/\overline{\nu}$, being $\overline{\nu}$
a constant such that inequality $\nu_i\leq \overline{\nu}$ holds at each spatial cell. Such limitation on $\Delta t$ ensures good accuracy but could lead to violation of the stability condition of the upwind scheme used in the free streaming sub-step.  To overcame this difficulty, we note that the exact solution of the streaming 
term is
\begin{equation}
\label{exact}
f(x,{\bf v},t+\Delta t) = f(x-v_{x}\Delta t,{\bf v},t)
\end{equation}
Thus, for each molecular velocity, $v_x(j_x)$, the value of the distribution 
function in the cell $(i,{\bf j})$ of the phase space can be obtained
by first translating the distribution function by a number of cells 
equal to the integer part of the Courant number 
$C=v_x(j_x)\Delta t/\Delta x$ , 
$\left[C\right]$,
and then applying expressions (\ref{eq:upwind}) for the residual time step advancement.
It should be observed that the density, bulk velocity and temperature obtained from
the discretized Maxwellian distribution function $\Phi_{i,\bj}$ are 
not exactly equal to $n_i$, $\bV_i$ and $T_i$. To ensure exact conservation of mass momentum and energy, the discretized $\Phi_{i,\bj}$ should be computed from Eq. (\ref{eq:Maxw}) by using
effective values $\tilde n_i$, $\tilde \bV_i$ and $\tilde T_i$ which are obtained by requiring
that the moments of the \emph{discretized} Maxwellian coincide with $n_i$, $\bV_i$ and $T_i$ \cite{f91}. The adoption of the correction method is not always necessary, since
mass, momentum and energy errors are very small, even for coarse velocity space grids.
Numerical tests have shown that calculating effective local values $\tilde n_i$, $\tilde \bV_i$ and $\tilde T_i$ to force exact conservation of collisional invariants did not affect appreciably
the solutions of the problems described below.\\
As is clear, both the free streaming and the relaxation sub-steps can be easily parallelized, each of them consisting of a number of independent threads.

\section{GPU and  $\mbox{CUDA}^{\mbox{\tiny TM}}$ overview}

$\mbox{NVIDIA}^\circledR$ 
GPU is built around a fully programmable processors array organized
into a number of multiprocessors with a SIMD-like architecture 
\cite{n08}, i.e. at any given clock cycle, each core of the multiprocessor 
executes the same instruction but operates on different data.
$\mbox{CUDA}^{\mbox{\tiny TM}}$ 
is the high level programming language 
specifically created for developing applications on this platform. \\
A $\mbox{CUDA}^{\mbox{\tiny TM}}$ 
program is organized into a serial program
which runs on the host CPU and one or more
kernels which define the computation to be performed in parallel by a massive 
number of threads. 
Threads are organized into a three-level hierarchy.
At the highest level, all threads form a grid; they all execute the same
kernel function. Each grid consists of many different blocks which contain
the same number of threads. 
A single multiprocessor can manage a number of blocks concurrently 
up to the resource limits. 
Blocks are independent, meaning that a kernel must 
execute correctly no matter the order in which blocks are run.
A multiprocessor executes a group of threads beloging to the active block,
called warp.
All threads of a warp execute the same instruction but operate on different 
data. 
If a kernel contains a branch and threads of the same warp follow different
paths, then the different paths are executed sequentially (warp divergence) 
and the total run time is the sum of all the branches. 
Divergence and reconvergence are managed in hardware but may have a serious 
impact on performance.
When the instruction has been executed, the multiprocessor
moves to another warp. In this manner 
the execution of threads is interleaved rather than simultaneous. \\
Each multiprocessor has a number of registers which are dynamically 
partitioned among the threads running on it. Registers are memory spaces 
that are readable and writable only by the thread to which they are 
assigned. Threads of a single block are allowed to syncronize with each other 
and are available to share data through a high-speed shared memory. 
Threads from different blocks in the same grid may coordinate only via 
operations in a slower global memory space which is readable and writeable 
by all threads in a kernel as well as by the host. 
Shared memory can be accessed by threads within a block as quickly as
accessing registers. On the contrary, I/O operations involving global memory are particularly
expensive, unless access is coalesced \cite{n08}.
Because of the interleaved warp execution,
memory access latency are partially hidden, i.e.,
threads which have read their data can be performing
computations while other warps running on the same multiprocessor are
waiting for their data to come in from global memory. 
Note, however, that GPU global memory is still ten time faster than the main
memory of recent CPUs.\\ 
Code optimization is a delicate task.
In general, applications which require many arithmetic operations between
memory read/write, and which minimize the number of out-of-order memory
access, tend to perform better. Number of blocks 
and number of threads per block have to be chosen carefully.
There should be at least as many blocks as multiprocessors in the
device. Running only one block per multiprocessor can force the multiprocessor 
to idle during thread synchronization and device memory reads.
By increasing the number of blocks, on the other hand, 
the amount of available shared memory for each block diminuishes.
Allocating more threads per block is better for efficient time slicing,
but the more threads per block, the fewer registers are available per thread.

\section{Shock wave}\label{sec:shockwave}
\subsection{Formulation of the problem}
The propagation of a planar shock wave is a classical application of kinetic 
equations which is a rather natural choice as a benchmark problem because of 
the considerable number of previous studies \cite{lnc62,kprw08}. 
In the wave front reference frame, the stationary flow field is assumed to be governed by the one-dimensional steady BGKW equation
\begin{equation}
\label{BGK}
v_{x} \frac{\partial f}{\partial x} =
\nu (\Phi-f)
\end{equation}
$x$ being the spatial coordinate which spans the direction normal to the 
(planar) wave front.
It is further assumed that, far from the wave front, the distribution function $f(x,\bv)$ satisfies the boundary conditions
\begin{equation}
 \lim_{x\rightarrow \mp \infty}f(x,\bv)=\Phi^{\mp}(\bv)=\frac{n^{\mp}}{\left( 2\pi R T^{\mp}\right)^{3/2}}\exp\left[-\frac{(v_x-V^{\mp})^2+v_y^2+v_z^2}{2R T^{\mp}} \right]
\label{eq:bc_shock}
\end{equation} 
where $n^{\mp}$, $V^{\mp}$ and $T^{\mp}$ are the upstream and downstream values of number density, velocity and temperature, respectively. The parameters
of the equilibrium states specified by Eq. (\ref{eq:bc_shock}) are connected by the Rankine-Hugoniot
relationships
\begin{equation}
\label{RH}
\frac{V^{-}}{V^{+}}=\frac{n^{+}}{n^{-}}=\frac{4(M^{-})^{2}}{(M^{-})^{2}+3}
\mbox{\hspace{1cm}}
\frac{T^{+}}{T^{-}}=\frac{\left[5(M^{-})^{2}-1\right]
                          \left[(M^{-})^{2}+3\right]}{16(M^{-})^{2}}
\end{equation}
In Eqs. (\ref{RH}) $M^{-}$ denotes the upstream infinity Mach number 
defined as
 \begin{equation}
M^{-}= \frac{V^{-}}{\left(\gamma RT^{-}\right)^{1/2}}
\end{equation}
being $\gamma=5/3$ the specific heat ratio of a monatomic gas.\\
The numerical scheme described in section \ref{sec:TBN} has been adopted to obtain approximate solutions of Eq. (\ref{BGK}) with boundary conditions (\ref{eq:bc_shock}) as long time limit of solutions of Eq. (\ref{eq:1D}) with identical boundary conditions and initial condition
\begin{equation}
f(x,\bv|0)=
\begin{cases}
\Phi^{-}(\bv) & x<0\\
\Phi^{+}(\bv) & x>0
\end{cases}
\end{equation}
The computations reported have been carried out for both a weak,
$M^{-}=1.5$, and a medium, $M^{-}=3.0$, shock wave.
The collision frequency has been obtained from Eq. (\ref{eq:frequency}), assuming that
the viscosity is given by the expression
\begin{equation}
\mu(T)=\mu_0\left(\frac{T}{T_0}\right)^{0.74} 
\label{eq:viscosity}
\end{equation}
In Eq. (\ref{eq:viscosity}), $T_0$ is a reference temperature, $\mu_0$ is the value of the viscosity at the reference temperature. The temperature exponent has been set equal to 0.74 
to match the computational conditions of Ref. \cite{lnc62} whose results have been used to asses the accuracy of the calculations presented here. 
An adimensional form of the Eq. (\ref{eq:1D}) has been adopted in actual computations by
normalizing velocity $\bv$ to $\sqrt{2RT^{-}}$, time $t$ to $\tau^{-}=1/\nu^{-}$ and spatial coordinate $x$ to the mean free path
 $\lambda^{-}=\sqrt{2RT^{-}}\tau^{-}$. The reference value $\nu^{-}$ for the collision frequency has been obtained by setting $T_0=T^{-}$ in Eq. (\ref{eq:viscosity}).
The infinite physical space has been replaced by the finite interval $[-L/2,L/2]$ which has been
divided into $N_{x}$ identical cells of width $\Delta x=L/N_x$.
The adimensional size $L$ of the spatial domain has been set equal to $70$, varying $N_x$ between
$128$ and $18432$. The cell number $N_x$ has been increased well above the limit imposed by accuracy in order to investigate the GPU performances as a function of computational load.
Similarly, the velocity space has been replaced by a parallelepiped in which each velocity component
$v_{\alpha}$ varies in a finite interval, divided into $N_{v_\alpha}$ equal cells.
The position of parallelepiped in the velocity space and the cell number vary with the chosen Mach number. In case of the weak shock wave, $M^{-}=1.5$, the same number of grid points has been used for the three normalized velocity components by setting $N_{v_\alpha}=16$, with $v_{x}\in[-5,7]$ and $v_{y},v_{z}\in[-6,6]$. In case of the $M_{\infty}^{-}=3.0$ shock wave, the grid point setting has been changed to $N_{v_\alpha}=30$, with $v_{x}\in[-10,12]$ and $v_{y},v_{z}\in[-11,11]$. 
Finally, the normalized time step $\Delta t$ has been set equal to $0.05$.
Before describing the algorithm implementation and describing the results, it is worth observing that, for one and two-dimensional problems, the dimensionality of the velocity space associated with kinetic model equations having the structure of Eq. (\ref{eq:BGKW}) can be
accordingly reduced to one and two, respectively \cite{Chu1965}. The reduction has not been made in the present work to keep the general structure of a three-dimensional code and, as mentioned above, to investigate the hardware response to heavy computer storage demand.  
\subsection{$\mbox{CUDA}^{\mbox{\tiny TM}}$ implementation}
The code to numerically solve Eq. (\ref{BGK})
is organized into a host program, which deals with 
all memory management and other setup tasks, and two kernels running on the 
GPU. One performs the streaming step and the other one performs the collision
step and compute the macroscopic quantities as well.
Alghorithms \ref{alg_streaming} and \ref{alg_collision} list the pseudocodes
of both kernels. For clarity of presentation, the pseudocode of the streaming 
step refers to the case of $v_{x}>0$.
Because of their different impact on the code performance, we 
distinguish the slow global memory reads, $\Leftarrow$, and writes,
$\Rightarrow$, from the fast reads, $\leftarrow$, and writes, $\rightarrow$, 
from local registers and shared memory.

\noindent
As shown by Eq. (\ref{exact}), for each given cell of the velocity space,
the streaming step involves the distribution function evaluated 
at different space locations.
The key performance enhancing strategy is to allow threads to cooperate
in the shared memory. The threads should thus be grouped into as
many blocks as the cells in the velocity space with a number
of threads per block equals to the number of cells in the physical space.
In practical applications, however,
the number of cells in the physical space is
greater than the maximum allowable number of threads per block. 
In order to fit into the device's resources, hence, the number of threads 
per block, $N_{t}$, is set to a lower value which is
chosen to maximize the utilization of registers and shared memory usage. 
When a block become active, each thread loads one element of
the distribution function from global memory, stores it into shared 
memory (line 7)
and then update its value according to Eqs. (\ref{eq:upwind})
(line 13). This
procedure is then repeated sequentially $N_{x}/N_{t}$ times. 
To ensure non-overlapping access, threads are synchronized at the onset of 
both reading from and writing to the global memory (lines 12 and 15). 
In order the access to the global memory to be coalesced, the discretized 
distribution function has been organized such that the value which refers to 
cells which are adjacent in the physical space are stored in contiguous memory
locations. A random memory access would determine otherwise a performance 
bottleneck.
Threads which update boundary points perform calculations which are slightly 
different to account for the incoming Maxwellian flux from the
boundary of the domain (line 5). 
This leads to a thread divergence which determines some code inefficiency.
However, testing shows that the performance loss is small.

\noindent
According to Eq. (\ref{relaxation}), in order to perform 
the relaxation step in a cell of the phase space, no
information from nearby cells is needed. Hence the concurrent 
computation of the collision operator may be possible mapping each thread to a 
single cell in the phase space.
This choice naturally fits for GPUs. 
In order to evaluate the local Maxwellian, $\Phi$, 
however, one must first calculate in each cell of the physical space the 
macroscopic quantities, Eqs. (\ref{RH}). 
In the attempt of reducing data transfers 
from and to the global memory, the computation of the macroscopic quantities
(lines 1-9) and the collision step (lines 11-13) are then performed in the same
kernel, by having a thread associated to each cell of the physical space.
Although this choice reduce the overall number of threads, it is not quite
limiting since for realistic three-dimensional problems, one would probably 
refine the physical grid more than the velocity grid. 

\subsection{Results and discussion}
In this section, we first validate the code by solving  
the plane shock structure problem and then we evaluate its performance by
comparing the GPU and CPU execution times. 
We chose representative commercial products from both the CPU and GPU markets: 
$\mbox{Intel}^\circledR$ 
$\mbox{Core}^{\mbox{\tiny TM}}$ Duo Quad Q9300  
running at 2.50 GHz with 6 MB of L2 cache and with 
4 GB of main memory, and an 
$\mbox{NVIDIA}^\circledR$ GeForce GTX 260 with 
$\mbox{CUDA}^{\mbox{\tiny TM}}$ version 2.0. 
The GTX 260 consists of 24 streaming multiprocessors. Each multiprocessor
has 8 streaming processors for a total of 192 units, clocked at 1.24 GHz.
Each group of  streaming processors shares one 16 kB of fast per-block
shared memory while the GPU has 896 MB of device memory.

\noindent
Figures \ref{Mach}a and \ref{Mach}b show the velocity and
temperature profiles versus the $x$ coordinate.
Solid and dashed lines are the results from the
numerical solution of Eq. (\ref{BGK}) for $M^{-}=1.5$ and 
$M^{-}=3$, respectively. Solid circles and squares are the results
presented in Ref. \cite{lnc62} for $M^{-}=1.5$ and 
$M^{-}=3$, respectively. 
The agreement is good and provides a validation of the numerical code.

\noindent
The performance of the GPU implementation is compared against the
single-threaded version running on the CPU by computing the
speedup factor $S=T_{\mbox{\tiny CPU}}/T_{\mbox{\tiny GPU}}$,
where $T_{\mbox{\tiny CPU}}$ and $T_{\mbox{\tiny GPU}}$
are the times used by the CPU and GPU to process at each time step one element 
of the discretized distribution function, respectively. 
We first examine the performance of each kernel and
then analyze the overall speedup of the program. 
Times are measured after initial setup, e.g., after file I/O,
and do not include the time required to transfer data between
the disjoint CPU and GPU memory spaces.

\noindent
Figure \ref{speed_cv} shows the speedup of the streaming kernel, 
$S_{\mbox{\tiny s}}$,
versus the number of cells in the physical space. Solid line with circles
and dashed line with squares are the results for a different number of
cells in the velocity space, 
$N_{v_{\alpha}}=16$ and $N_{v_{\alpha}}=30$, respectively.
In both cases, the speedup sharply increases and then level off at about 
$N_{x} \backsimeq 3000$, where the GPU capability is fully exploited.
For a greater number of cells in the physical space,  
the streaming step scales linearly with the
elements of the discretized distribution function.
The speedup decreases with the number of cells in the velocity space 
but, even in the worst case, it is still about $200$. \\
Figure \ref{speed_cl} shows the same as Fig. \ref{speed_cv} but for
the collision kernel, 
Unlike the streaming kernel, the speedup of the collision kernel, 
$S_{\mbox{\tiny c}}$, is
greatly improved by increasing the number of cells in the velocity space. \\
Figures \ref{time}a and \ref{time}b show the relative time,
spent to perform the streaming kernel, $T_{\mbox{\tiny s}}$, and 
the collision kernel, $T_{\mbox{\tiny c}}$,
versus the number of cells in the physical 
space, for $M^{-}=1.5$ and $M^{-}=3$, respectively.
As expected, the collision is much more time consuming than the
streaming kernel. Moreover, the relative time does not appear to depend
strongly on the number of cells in the velocity space.  \\
Figure \ref{speed_tot} shows the overall speedup, $S_{\mbox{\tiny t}}$,
for the two test cases. Notation is the same as Figs. \ref{speed_cv} and
\ref{speed_cl}. The overall speedup for each case is in between 
$S_{\mbox{\tiny s}}$ and $S_{\mbox{\tiny c}}$, which is not unexpected. 
In fact, $S_{\mbox{\tiny t}}$ is
the weighted average of the streaming and collision speedups with weight the 
relative time $T_i$ spent by the GPU to execute each kernel, i.e.,
$ S_{\mbox{\tiny t}} = T_{\mbox{\tiny s}} S_{\mbox{\tiny s}} + 
                       T_{\mbox{\tiny c}} S_{\mbox{\tiny c}}$.

\section{Driven cavity}

\subsection{Formulation of the problem}
The driven cavity flow is a classical multidimensional benchmark problem 
since, in spite of its simple geometry, it contains most of the features
which appear in more complicated problems described by kinetic equations.  
A gas is confined in a two-dimensional square cavity and
the flow is driven by a uniform translation of the
top with velocity $U_{\mbox{\tiny W}} {\bf e}_{x}$. 
The gas flow is supposed to 
be governed by the two-dimensional steady BGKW equation

\begin{equation}
\label{BGK_cavity}
v_{x} \frac{\partial f}{\partial x} +
v_{y} \frac{\partial f}{\partial y} =
\nu (\Phi-f)
\end{equation}
It is further assumed that all the walls are isothermal and that the particles 
which strike the walls are re-emitted according to the Maxwell's scattering 
kernel with complete accommodation

\begin{equation}
\label{eq:bc_cavity}
f(\bx,\bv) = \Phi_{\mbox{\tiny W}}({\bf v}) =
\frac{n_{\mbox{\tiny W}}}{(2\pi RT_{0})^{1/2}}
\exp{\left[ -\frac{(\bv-\bV_{\mbox{\tiny W}})^{2}}
                  {2RT_{0}} \right]}, 
\quad (\bv-\bV_{\mbox{\tiny W}})\circ \bn>0
\end{equation} 
where ${\bx}$ is a point of the boundary, $\bn$ the inward normal,  ${\bf V}_{\mbox{\tiny W}}$
the wall velocity and $n_{\mbox{\tiny W}}$ the wall density defined as
\begin{equation}
n_{\mbox{\tiny W}} = \left( \frac{2\pi}{RT_{0}} \right)^{1/2}
\int_{({\bv}-{\bV}_{\mbox{\tiny W}})\circ {\bn}<0}
|({\bv}-{\bV}_{\mbox{\tiny W}})\circ {\bn}| f \; d{\bv}
\end{equation}
in order to impose zero net mass flux at any boundary point.\\
The two-dimensional extension of the numerical scheme described in section 
\ref{sec:TBN} has been adopted to obtain approximate solutions of Eq. 
(\ref{BGK_cavity}) with boundary conditions (\ref{eq:bc_cavity}) as long
time limit of the unsteady problem.
The adimensional form of the governing equation has been obtained as described in section
\ref{sec:shockwave}.
In Ref. \cite{VVS08}, the cavity flow problem has been solved by assuming
that $V_{\mbox{\tiny W}} \ll \sqrt{2RT_{0}}$ and thus 
Eq. (\ref{BGK_cavity}) has been linearized around the equilibrium state at 
rest. In order to reproduce these results, the dimensionless lid velocity
is here set to $0.01$. The gas is thus in a weakly non-equilibrium state
and the nonlinear results approach the linearized ones.  
The square cavity, $[0,\delta]\times[0,\delta]$, has been divided into 
$N_{x}=N_{y}=160$ cells with uniform width. Here $\delta$ is the rarefaction 
parameter which is proportional to the inverse of the Knudsen number.
The cavity flow problem has been solved over a wide range of the rarefaction 
parameter, $\delta \in [0.1-10]$. 
The computational grid in the physical space has been chosen to achieve
the convergence of the results in the whole range of rarefaction parameter 
considered.  
The number of velocity cells have been set $N_{v_{\alpha}}=20$
with $v_{x},v_{y},v_{z} \in [-3,3]$. 
Finally, the time step has been varied in the range 
$10^{-4}-10^{-2}$ depending on the rarefaction parameter.

\subsection{Results and discussion}
Figures \ref{profili}a and \ref{profili}b
show the profiles of the horizontal component of 
the velocity, $V_{x}/V_{\mbox{\tiny W}}$, on the vertical plane crossing the center of a
square cavity and the vertical component of the velocity, $V_{y}/V_{\mbox{\tiny W}}$, 
on the horizontal plane crossing the center of the top vortex, respectively, 
for two different value of the rarefaction parameter, $\delta=0.1,10$. 
Solid lines are the numerical results obtained by solving 
Eq. (\ref{BGK_cavity}) with the parallel code, 
solid circles are the results reported in 
Ref. \cite{VVS08}. The agreement is quite good.
The near linear profiles of the velocity in the central core of the cavity
indicate the uniform vorticity region. 
In order to proceed with a more detailed comparison, we introduce two overall
quantities, namely the mean dimensionless shear stress, $D$, along the moving 
plate and the dimensionless flow rate, $G$, of the main vortex.
The former quantities is obtained by integrating the shear stress along the lid 
of the cavity, the latter by integrating the x-component of the velocity
profile along the plane crossing the center of the cavity from the center
of the top vortex up to the lid.
Table \ref{table} compares the prediction of $D$ and $G$ obtained 
by solving Eq. (\ref{BGK_cavity}) with the parallel code and the  
values reported in Ref. \cite{VVS08}, for different values of 
the rarefaction parameter. The agreement is good, the greatest mismatch
being for the drag coefficient at $\delta=10$. 
However the discrepancy is easily removed by increasing the number 
of cells in the physical space.
Figure \ref{tempi_fase} shows the time spent for processing one cell of the 
the phase space at each time step, expressed in nanoseconds, versus the number 
of cells used to discretize the physical space, $N_{r}=N_{x}$ for the
one-dimensional code and $N_{r}=N_{x}+N_{y}$ for the two-dimensional code.
The solid and dashed lines are the results for the shock wave and driven 
cavity flow problems, respectively. The one-dimensional and two-dimensional
codes fully utilize the GPU when the number of cells in the physical space 
is about $6000$ and $9000$, respectively. This difference is due to the fact
that the one-dimensional code makes a better use of GPU's registers and shared 
memory. For a greater number of cells, the total execution time increases 
linearly and hence the time spent for processing one cell of the phase space 
at each time step is nearly constant. 
When the GPU is fully exploited, the two-dimensional code is
slower than the one-dimensional code by a factor of about $2$, which is not 
unexpected because of thread divergence determined by the greater number of 
boundary cells.
Although the sequential version of the two-dimensional code has not
been written, it can be safely inferred that the speedup of the two-dimensional code is about an half of the speedup of the one-dimensional code, on the basis of these results.

\section{Conclusions}
The aim of this paper is to describe the development of an algorithm 
to solve kinetic equations by exploiting the computing power of modern GPUs.
Test gas flows have been studied by adopting the Bhatnagar-Gross-Krook-Welander (BGKW) 
kinetic model for the collision term in combination with a simple finite difference scheme. 
Numerical experiments with the one-dimensional shock wave structure problem 
and the two-dimensional driven cavity flow indicate that it is possible to cut 
down the computing time of the sequential codes up to two order of magnitudes. 
For instance, the solution of the two-dimensional unsteady driven cavity
flow for $\delta=10$ with $N_{v_{\alpha}}=20$, $N_{x}=N_{y}=160$ 
took about $7$ minutes to execute $4100$ time steps. 
It is worth to notice that if the specific two-dimensional nature of the problem 
is taken into account, then the dimensionality of the velocity space can be reduced
by a standard projection procedure \cite{Chu1965}. In this case, the computing time can be further
reduced to about $40$ seconds while keeping the same accuracy level.
The algorithm described can easily be extended to three dimensions and
to non-equilibrium flows involving mixtures, and/or chemical reactions 
\cite{gast08}.
Extension to polyatomic gas is possible as well provided that one
adopts a proper representation of the distribution function to cope with 
the enlargement of the phase space due to internal degrees of freedom \cite{f07}.
This paper describes the first stage in the development of algorithms for 
solving non-equilibrium gas flows onto GPUs. The extension of this algorithm 
to semi-regular method of solution to the full Boltzmann equation is presently 
being investigated and it will be considered in a future paper.

\section*{Acknowledgment}
Support received from {\bf Fondazione Cariplo}
within the framework of project \emph{``Fenomeni dissipativi e di rottura in
micro e nano sistemi elettromeccanici''}, and
{\bf Galileo Programme} of Universit\`a Italo-Francese within
the framework of project MONUMENT (MOdellizzazione NUmerica in MEms e
NanoTecnologie) is gratefully acknowledged.
The authors wish to thank Professor Dimitris Valougeorgis for providing his
numerical results.

\newpage

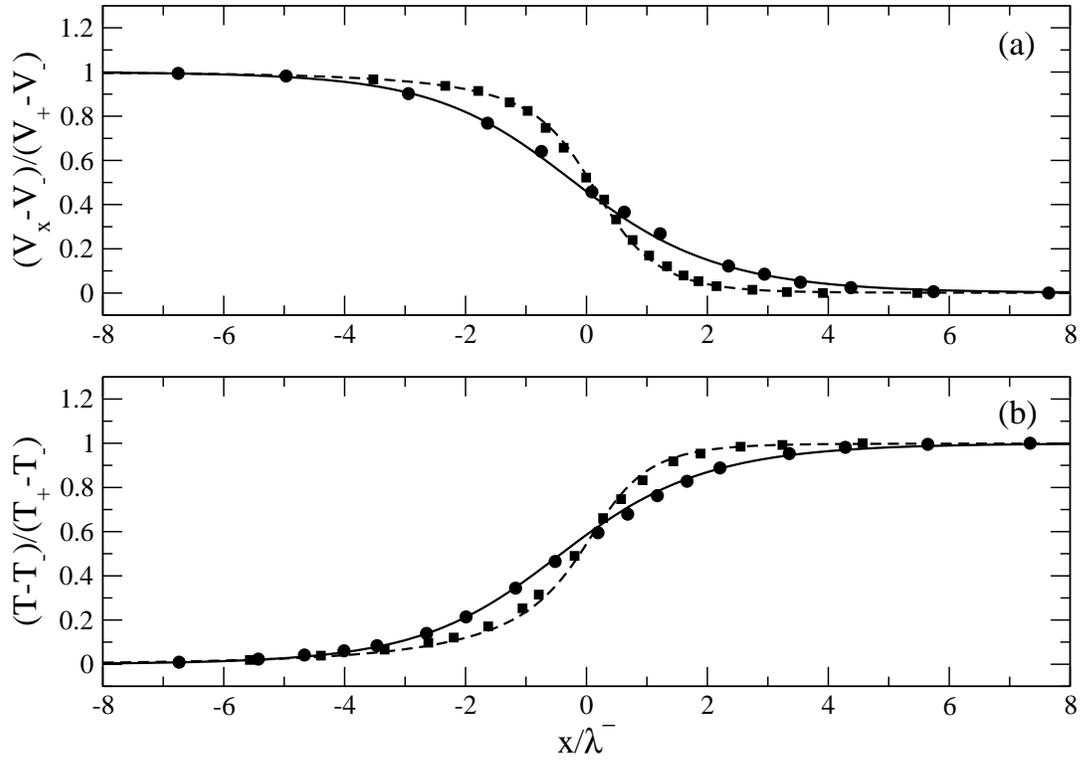
\begin{figure}[h]
\begin{center}
  \epsfig{file=Mach.eps,height=10cm}
  \caption{}
  \label{Mach}
\end{center}
\end{figure}

\noindent
Captions to Figures:

\vspace{1cm}

\noindent
Fig. \ref{Mach}: (a) mean velocity and (b) temperature profiles 
versus the $x$ coordinate. Solid and dashed lines are the results obtained
with the parallel code for $M^{-}=1.5$ and 
$M^{-}=3$, respectively. Solid cirlces and squares are the results
presented in Ref. [20] for $M^{-}=1.5$ and 
$M^{-}=3$, respectively. 

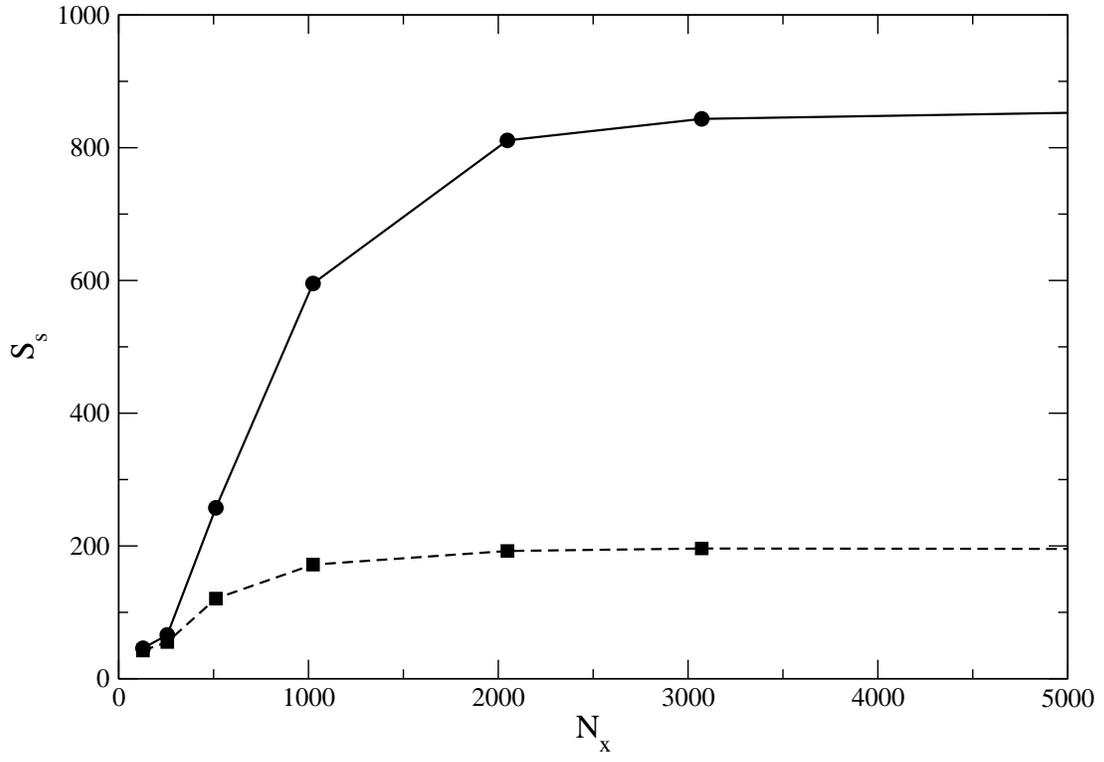
\begin{figure}[h]
\begin{center}
  \epsfig{file=speed_cv.eps,height=10cm}
  \caption{}
  \label{speed_cv}
\end{center}
\end{figure}

\noindent
Fig. \ref{speed_cv}: speedup of the streaming kernel, $S_{\mbox{\tiny s}}$, 
versus the number of cells in the physical space, $N_{x}$.
Solid line with circles: $N_{v_{\alpha}}=16$; 
dashed line with squares: $N_{v_{\alpha}}=30$. 

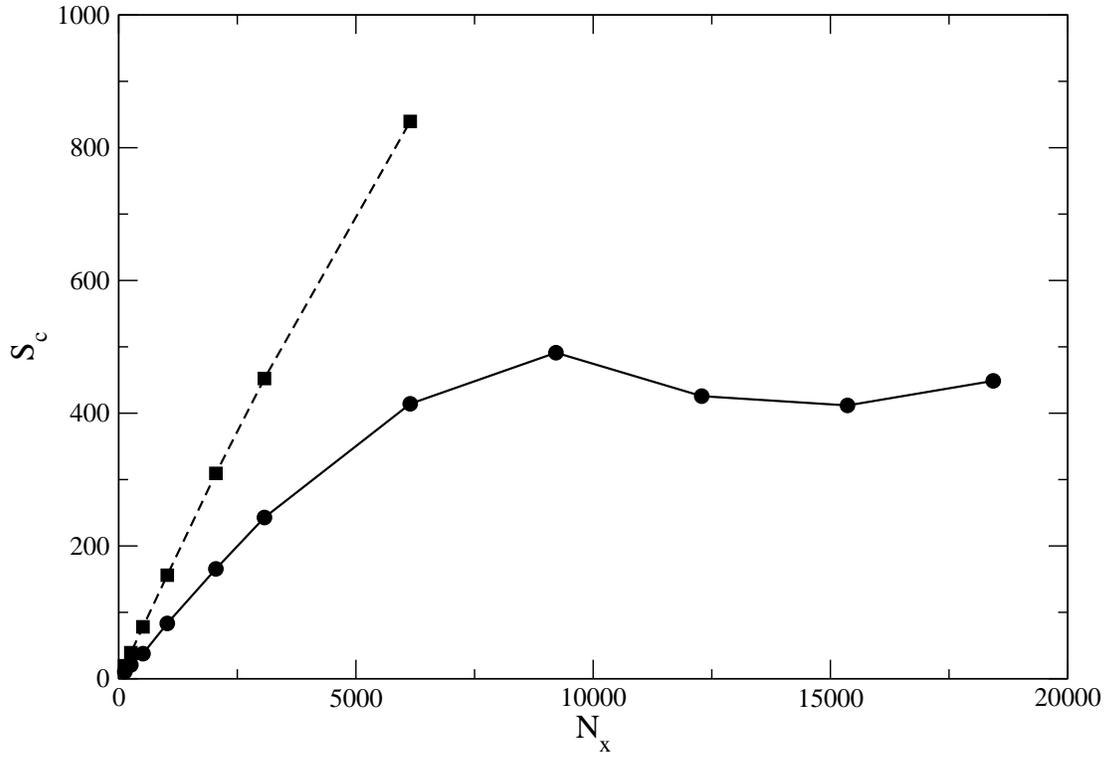
\begin{figure}[h]
\begin{center}
  \epsfig{file=speed_cl.eps,height=10cm}
  \caption{}
  \label{speed_cl}
\end{center}
\end{figure}

\noindent
Fig. \ref{speed_cl}: speedup of the collision kernel, $S_{\mbox{\tiny c}}$, 
versus the number of cells in the physical space, $N_{x}$. 
Solid line with circles: $N_{v_{\alpha}}=16$; 
dashed line with squares: $N_{v_{\alpha}}=30$. 

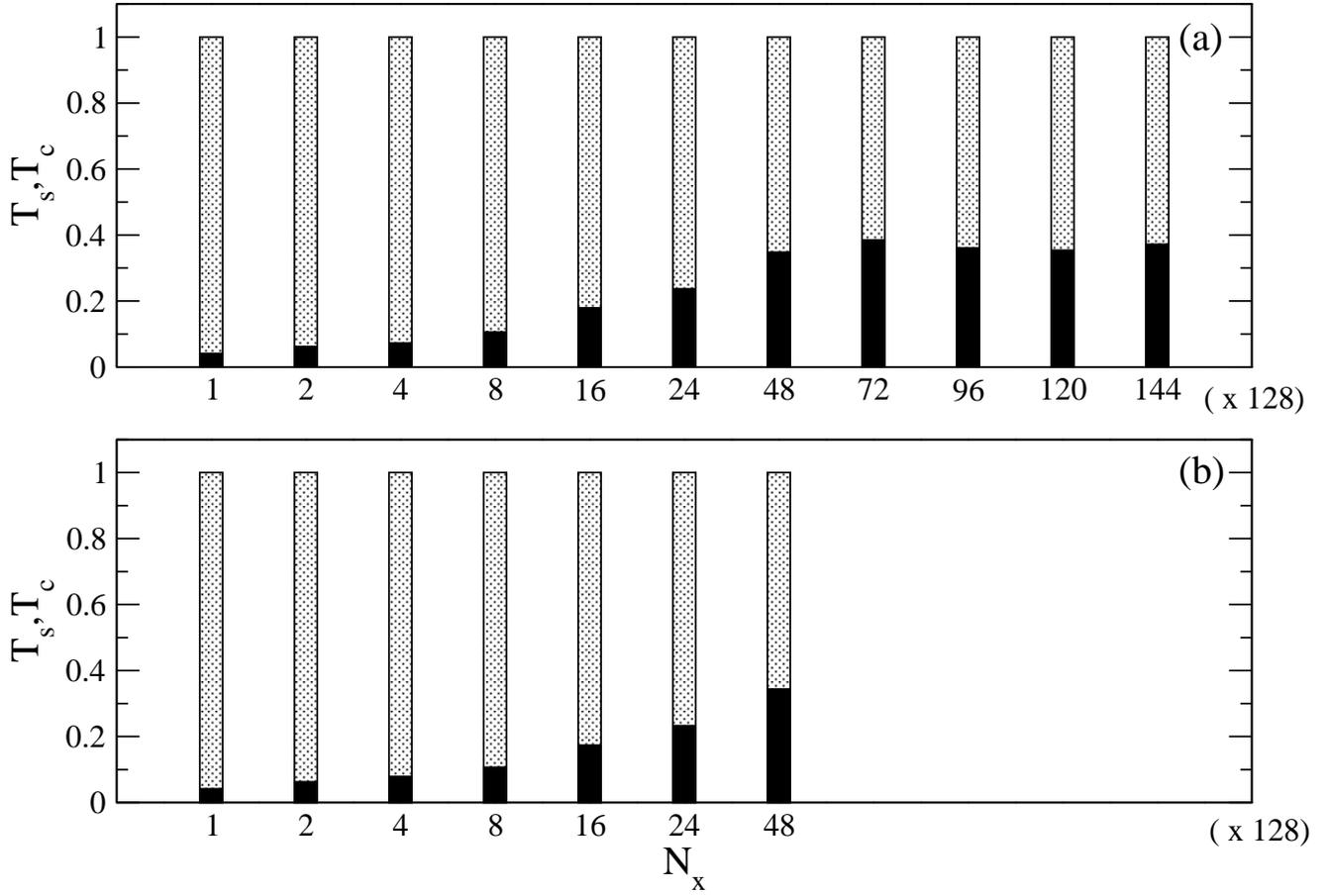
\begin{figure}[h]
\begin{center}
  \epsfig{file=time.eps,height=12cm}
  \caption{}
  \label{time}
\end{center}
\end{figure}

\noindent
Fig. \ref{time}: relative time spent on the streaming and collision kernel
for (a) $N_{v_{\alpha}}=16$ and (b) $N_{v_{\alpha}}=30$. 
Solid bar: streaming kernel; pattern bar: collision kernel.

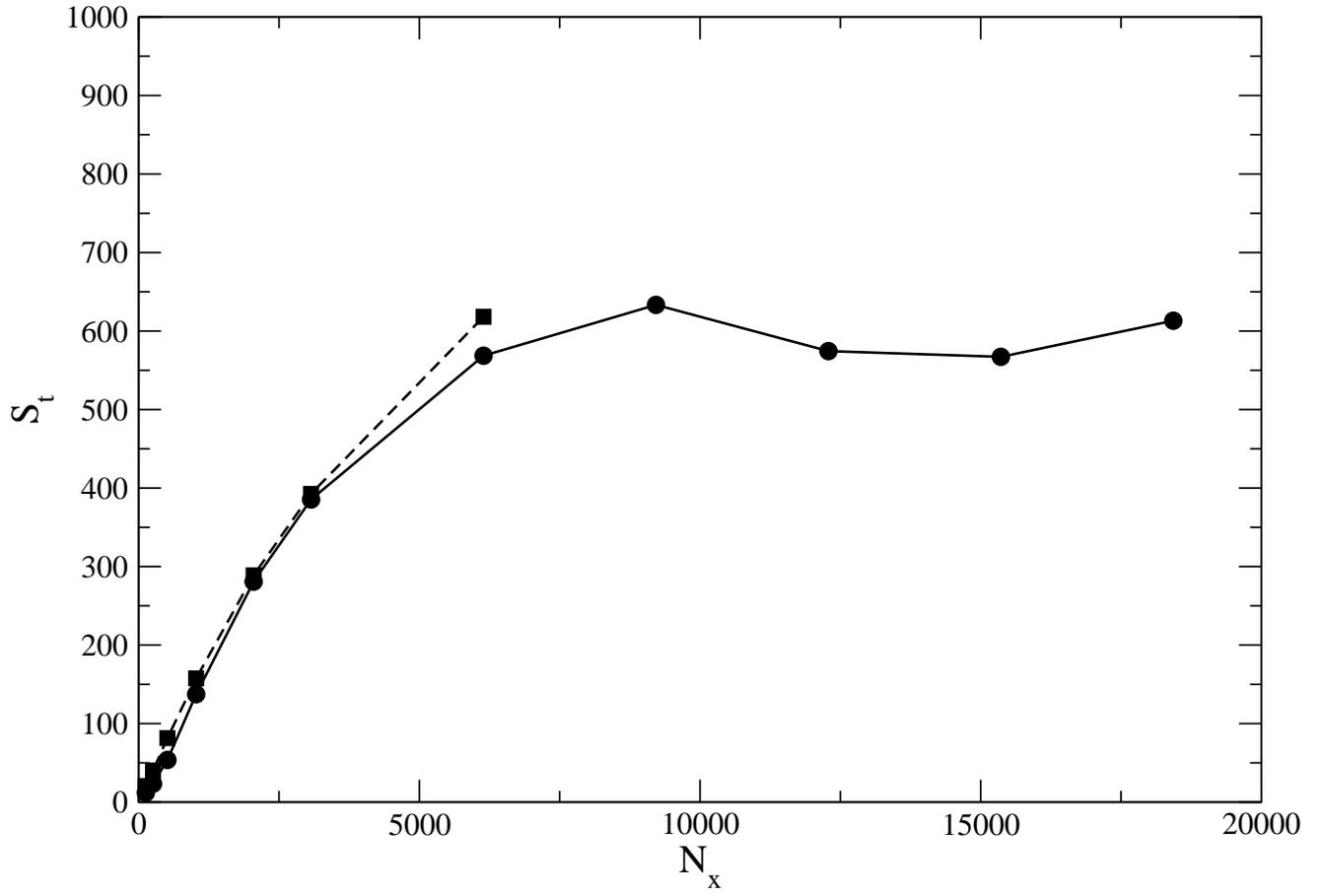
\begin{figure}[h]
\begin{center}
  \epsfig{file=speed_tot.eps,height=12cm}
  \caption{}
  \label{speed_tot}
\end{center}
\end{figure}

\noindent
Fig. \ref{speed_tot}: overall speedup, $S_{\mbox{\tiny tot}}$, 
versus the number of cells in the physical space, $N_{x}$.
Solid line with circles: $N_{v_{\alpha}}=16$; 
dashed line with squares: $N_{v_{\alpha}}=30$. 

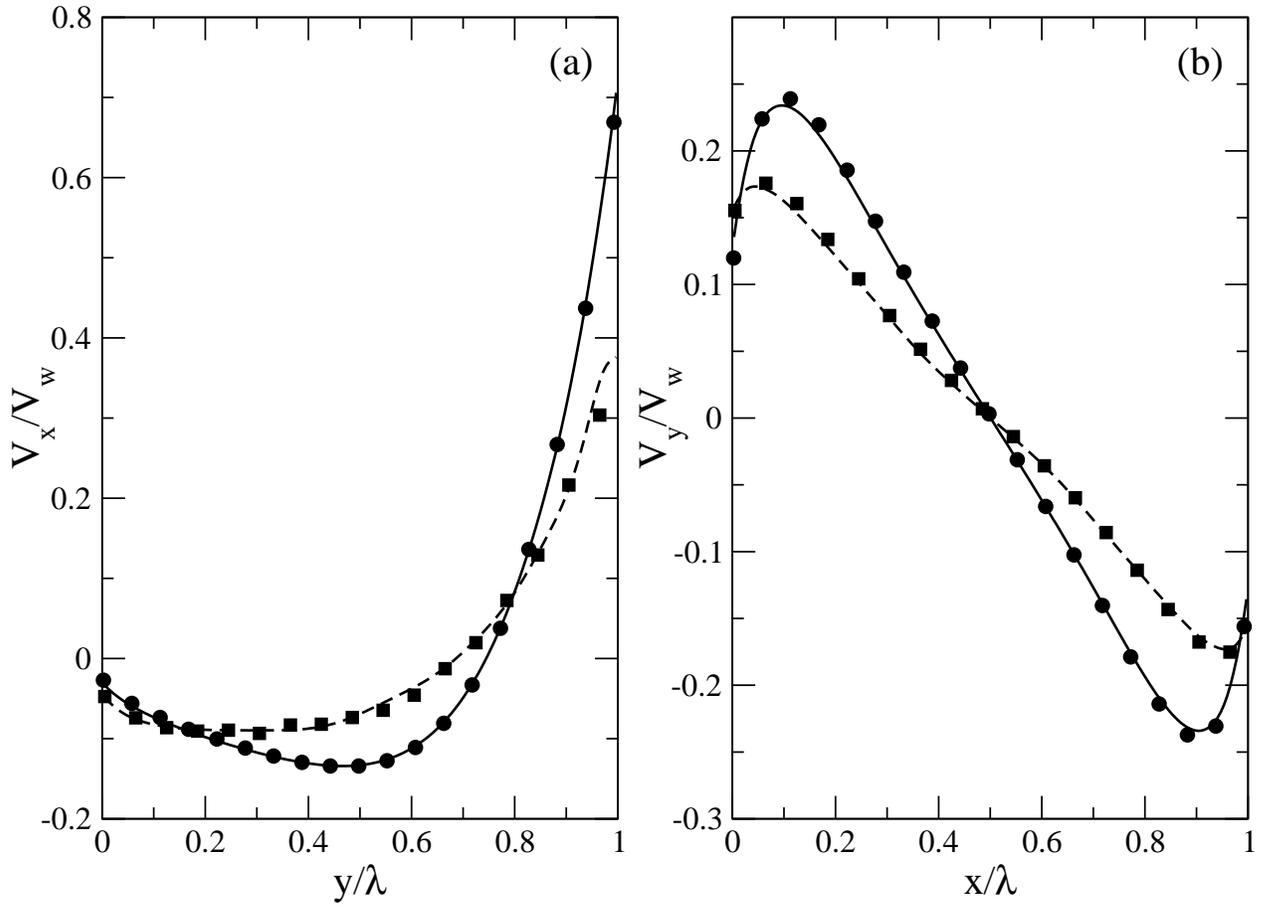
\begin{figure}[h]
\begin{center}
  \epsfig{file=profili.eps,height=12cm}
  \caption{}
  \label{profili}
\end{center}
\end{figure}

\noindent
Fig. \ref{profili}: profiles of (a) the horizontal component of the velocity 
on the vertical plane crossing the center of the cavity and (b) the vertical
component of the velocity on the horizontal plane crossing the center
of the top vortex. Solid lines: numerical solution obtained with the 
parallel code; solid circles: solution reported in Ref. \cite{VVS08}. 

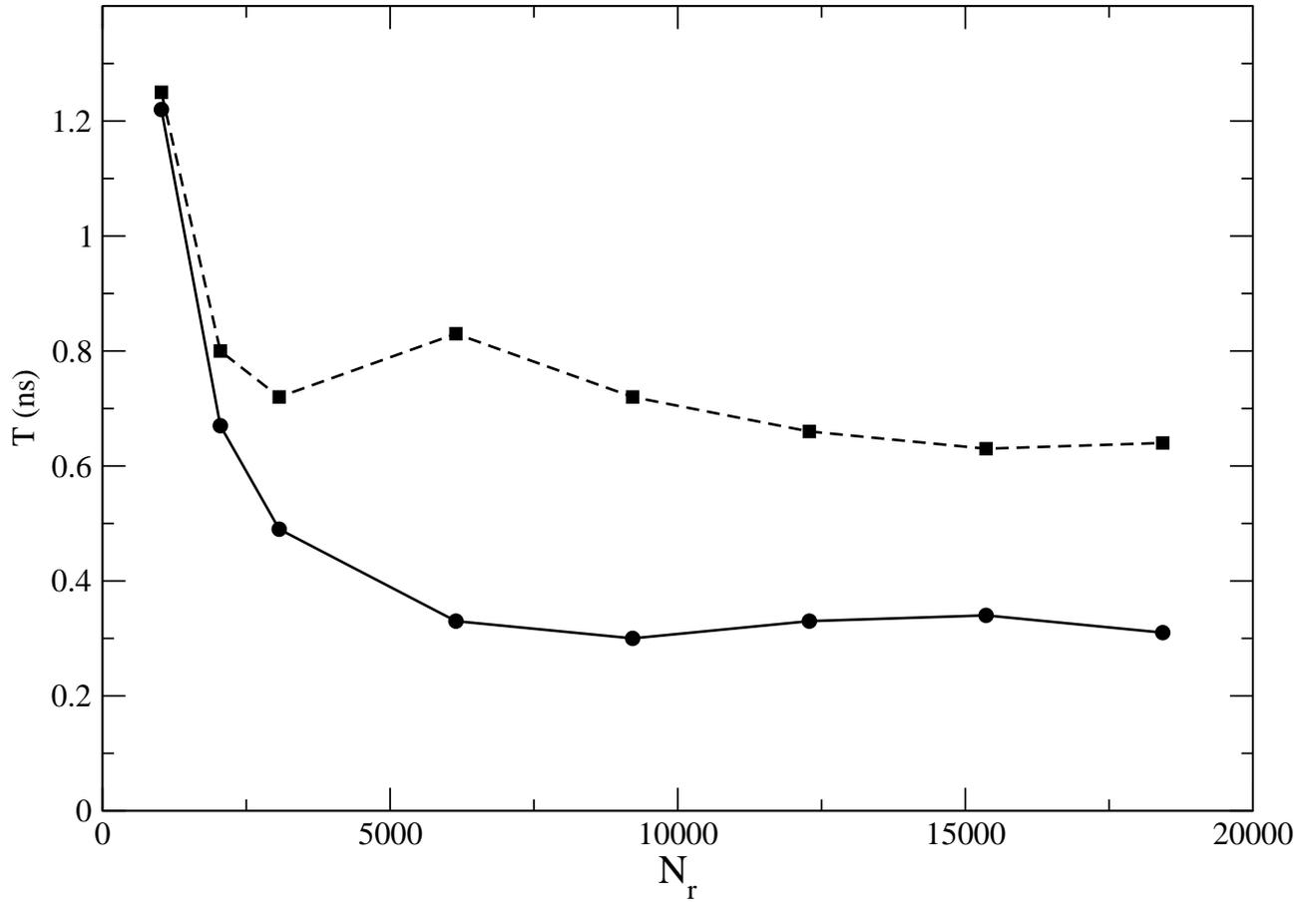
\begin{figure}[h]
\begin{center}
  \epsfig{file=tempi_fase.eps,height=12cm}
  \caption{}
  \label{tempi_fase}
\end{center}
\end{figure}

\noindent
Fig. \ref{tempi_fase}: time spent for processing one cell of the phase 
space versus the number cells used to discretized the physical space,
$N_{r}$. Solid line: one-dimensional code. Dashed line: two-dimensional code. 
$N_{v_{\alpha}}=16$.

\noindent
Table \ref{table}: drag coefficient, $D$, and reduced flow rate, $G$, 
versus the rarefaction parameter, $\delta$.

\newpage

\begin{algorithm}
\caption{GPU pseudo-code of the streaming step}
\label{alg_streaming}
\begin{algorithmic}[1]
\REQUIRE $\left[C\right]$ 
         is the integer part of the Courant number\\
\REQUIRE $\tilde{C}$ 
         is the fractional part of the Courant number\\
\REQUIRE $N_{x}$ is the number of cells in the physical domain \\
\REQUIRE $N_{t}$ is the number of threads in each block\\
\REQUIRE $t=0,\ldots N_{t}-1$ is the index of the thread within the block\\
\STATE $r \leftarrow  N_{x}-N_{t} -\left[C\right]$ \\
\STATE $w \leftarrow N_{x} -N_{t}$ \\
\FOR{$i=1 \; \; \mbox{to} \; \; N_{x}/N_{t}$}
\IF{$r+t<0$}
\STATE $f_{\mathrm{sh}}(t+1) \leftarrow \Phi^{-}_{\bf j}$
\ELSE
\STATE $f_{\mathrm{sh}}(t+1) \Leftarrow f^{n}_{r+t,\bf{j}}$
\ENDIF
\IF{$t = 0$} 
\STATE $f_{\mathrm{sh}}(0) \Leftarrow f^{n}_{r-1,{\bf j}}$
\ENDIF
\STATE syncthreads
\STATE $f_{\mathrm{rg}}
       \leftarrow 
       (1-\tilde{C}) \; f_{\mathrm{sh}}(t+1) + 
       \tilde{C} \; f_{\mathrm{sh}}(t)$
\STATE $f_{\mathrm{rg}} \Rightarrow \tilde{f}^{n+1}_{w+t,{\bf j}}$
\STATE syncthreads 
\STATE $r \leftarrow r - N_{t}$
\STATE $w \leftarrow w - N_{t}$
\ENDFOR
\end{algorithmic}
\end{algorithm}

\newpage

\begin{algorithm}
\caption{GPU pseudocode of the collision step}
\label{alg_collision}
\begin{algorithmic}[1]
\REQUIRE $i$ is global index of the thread inside the grid\\
\FORALL{${\bf j}$}
\STATE $f_{\mathrm{rg}} \Leftarrow \tilde{f}^{n+1}_{i,{\bf j}}$
\STATE $n_{\mathrm{rg}} \leftarrow n_{\mathrm{rg}} + f_{\mathrm{rg}}$
\STATE ${\bf V}_{\mathrm{rg}} \leftarrow {\bf V}_{\mathrm{rg}} + 
                                         {\bf v}_{\bf j} \; f_{\mathrm{rg}}$
\STATE $e_{\mathrm{rg}} \leftarrow e_{\mathrm{rg}} +  
                                   |{\bf v}_{\bf j}|^{2}\;f_{\mathrm{rg}}$
\ENDFOR
\STATE $n_{\mathrm{rg}} \leftarrow n_{\mathrm{rg}} \; \Delta \mathcal{V}$
\STATE ${\bf V}_{\mathrm{rg}} \leftarrow 
        {\bf V}_{\mathrm{rg}} / n_{\mathrm{rg}}$
\STATE $T_{\mathrm{rg}} \leftarrow (e_{\mathrm{rg}}/n_{\mathrm{rg}} - 
                                   |{\bf V}_{\mathrm{rg}}|^{2})/3$
\FORALL{${\bf j}$}
\STATE $f_{\mathrm{rg}} \Leftarrow \tilde{f}^{n+1}_{i,{\bf j}}$
\STATE $f_{\mathrm{rg}} \leftarrow 
          \left[1-\exp(-\nu_i\Delta t)\right]\Phi_{i,\bj}+
          \exp(-\nu_i\Delta t) f_{\mathrm{rg}}$
\STATE $f_{\mathrm{rg}} \Rightarrow 
        f^{n+1}_{i,{\bf j}}$
\ENDFOR
\STATE $n_{\mathrm{rg}} \Rightarrow n_{i}$
\STATE ${\bf V}_{\mathrm{rg}} \Rightarrow {\bf V}_{i}$
\STATE $T_{\mathrm{rg}} \Rightarrow T_{i}$
\end{algorithmic}
\end{algorithm}

\newpage

\newpage

\newpage

\newpage

\newpage

\newpage

\newpage

\newpage

\begin{table}[h]
\begin{center}
\begin{tabular}{|c||c|c||c|c|} \hline 
$\delta$ & D & D (Ref \cite{VVS08}) & G & G (Ref \cite{VVS08}) \\ \hline
0.1 & 0.675 & 0.678-0.676 & 0.0975 & 0.0973-0.0976 \\ \hline
1   & 0.624 & 0.625-0.631 & 0.103 & 0.104-0.105 \\ \hline
10  & 0.393 & 0.412-0.415 & 0.143 & 0.145-0.145 \\
\hline 
\end{tabular}
\caption{}
\label{table}
\end{center}
\end{table}

\end{document}